# The Quest for Gravity Agent:
# from Newton to Einstein and Feynman

**Yurij V. Baryshev**

Astronomical Department of the Saint-Petersburg State University,
198504 St.Petersburg, Russia,

*"Amicus Plato, sed magis amica veritas"*   *Aristotle*

**The Problem of Gravity Quantization**

Contemporary discussions [1] of theoretical problems generated by classical (non-quantum) Black Hole event horizon and by gravitational wave energy transfer and localization, have demonstrated that gravity quantization is unavoidable forthcoming paradigm shift in gravity physics. The key question is to choose a way for gravity quantization – one approach is the space-time quantization, while another approach is quantization of material gravitation field in Minkowski space-time, as for other fundamental physical interactions.

From history of physics we know that Newton was the first who raised scientific question about nature of gravitation. In his third letter to Bentley, Newton wrote: "Gravity must be caused by an Agent acting constantly according to certain Laws; but whether this Agent be material or immaterial, I have left to the Consideration of my Readers" [2].

Surprisingly in the beginning of $21^{st}$ century this alternative in gravity understanding, to be "material" or "immaterial", is still under consideration by modern theoretical physics. Indeed, for Einstein's geometrical general relativity theory (GRT) [3] gravity is curvature of the "non-material" space-time, and in this sense, space-time presents the Newtonian "immaterial" agent of gravity. While for Feynman's relativistic quantum field theory (FGT) [4] gravity is a consequence of exchange by "material" gravitons – the quanta of gravitation field in Minkowski space-time.

This alternative opens gate where philosophical aspects of physics enter to gravitation theory at the level of basic initial principles, which determine interpretation of observational data and understanding gravity nature.

**Philosophy of Science on Geometry and Physics**

Modern geometrical paradigm for gravity understanding is based on the mathematical concept of metric space (space-time) and its curvature. Note that ontological status of space-time is principally different from matter, like properties of apple differ from apple itself.

If one asks a mathematician: what is geometry?, the reply will be unexpected [5]: "a part of mathematics is called geometry because a sufficient number of competent people think 'on emotional and traditional grounds' that it is a good name". This simply means that mathematics works with logical correctness of inferred formal theorems without testing it by physical experiments. However, for physics the relation of a theory to reality is the crucial problem, so *mathematical* concepts of coordinate, distance, and internal geometry are not equivalent to their *physical* counterparts.

Philosophy of physics considers important difference between mathematical abstract concepts (like numbers or forms) and physical quantities, which have physical dimensions (e.g. gram, centimeter, second). According to Bridgman's logic of physics [6] the essence of physical quantity is its measurability, i.e. its *relation to units* and measuring procedures.

This "first relativity" of physical quantities has important implication in the case of length measurement inside a curved space. Indeed, the physical distance between object $A$ and $B$ is not a number, but the number of unchangeable unit of length (meter stick), which can be defined only outside the curved space (free motion). For instance the unit of length in Euclidean space $E3$ induces the unchangeable unit of length in spherical subspace $S2$, embedded in $E3$. Having this in mind, in 1824 letter to German scholar Taurinus, famous mathematician and physicist Gauss wrote: "*I sometimes joke that it would be good, if Euclidean geometry were not true, because then we would have a priori absolute measure of length*". So the problem of absolute intrinsic units is connected with reality of embedding space in physics.

**Quantum Curved Space or Relativistic Quantum Field in Flat Space?**

The success of Einstein's *geometrical gravity* theory in explanation of classical relativistic gravity effects is generally recognized [7]. However there are some puzzling theoretical problems of the geometrical approach, such as the "energy problem" [8] (pseudo-tensor of energy-momentum) and the "information paradox" [1] (escaping information from classical event horizon). Also GRT is not a quantum theory and there are different attempts of quantization the curved Riemannian space-time (e.g. [9]) which have not tested yet. Moreover the concept of gravitation *energy quanta* cannot be properly defined in a theory where the local energy density does not exist.

Natural alternative approach is Feynman's *field gravity* theory [4], which consider material relativistic quantum field in Minkowski space, like in electrodynamics and quantum electrodynamics. Within FGT all basic physical concepts are common with other theories of fundamental physical interactions. So gravity force and localizable positive energy density of the gravitational field (static and free) exist inside and outside massive bodies [10]. Gravitational waves are generated by the source of energy-momentum and every radiated graviton carries away an amount of energy $h\nu$.

Up to now weak field gravity effects, which had been really tested by observations (including universality of free fall), cannot distinguish between Einstein's geometrical and Feynman's field (in flat Minkowski space) theories. However in the case of strong gravity predictions are essentially different [10].

According to Poincare [11], it is "convenient" to use in physics the flat Euclidean space, where Cartesian coordinates always exist and all physical laws can be formulated on the flat background. Modern theoretical physics gives fundamental reason for the flat "prior geometry": energy-momentum conservation laws directly follow from the maximal symmetry of Euclidean and Minkowski spaces (Noether's theorem). This also explains why in general relativity (which is based on curved space) there is famous "energy problem" – absence of localizable energy of gravity in GRT, which also is primary cause of the information paradox.

**Modern Observations of Astrophysical Black Holes and Gravitational Waves**

Observational relativistic astrophysics, which started in sixties of 20[th] century, now achieves its mature level with detection of gravitational waves and study strong gravity effects in relativistic compact objects (RCO).

Crucial new observational facts are: 1) the LIGO-Virgo detection of GW, i.e. transfer and localization of GW positive energy; 2) the EHT restriction on existence of the photon ring around BH in SgrA*; and 3) the Fe-line derivation of the accretion disk inner radius, which is less than Schwarzschild radius.

Gravitational-wave signals were detected by using Advanced LIGO and Virgo interferometric antennas [12]. This means that the positive gravitational field energy carried by gravitational waves, was localized by a GW detector, i.e. free gravitational field energy can be transformed to the kinetic energy of the moving LIGO mirrors. Hence the GW observations naturally point to the field nature of the gravitation interaction.

New possibility for observations close to horizon of supermassive BH candidates (BHC) comes from mm/submm wavelength VLBI Event Horizon Telescope (EHT) [13]. Event-horizon-scale structure in the supermassive black hole candidate at the Galactic Centre (SgrA*) can be achievable directly with submm EHT and this will give possibility to test relativistic and quantum gravity theories at the gravitational radius. The first results of EHT observations at 1.3mm surprisingly demonstrated that for the BHC in SgrA* there are no expected for BH the light ring with diameter $5.2R_{Sch}$ : observed BHC size $\theta_{obs} = 37$ μas, while GRT theoretical prediction for size of the ring $\theta_{ring} = 53$ μas. EHT observations have opened a new page in study of strong gravity effects in vicinity of BHCs. Note that in FGT there is no light ring because the gravity force is finite everywhere and RCO has no event horizon.

Surprising observational facts come also from X-ray studies of BHCs at centers of luminous Active Galactic Nuclei and stellar mass Black Hole Candidates [14]. Analysis of the iron K-line profiles and luminosity variability gave amazing result: the estimated radius of the inner edge ($R_{in}$) of the accretion disk around RCO is about $(1.2 − 1.4)$ Rg, where Rg $= GM/c^2 = R_{Sch}/2$, i.e. about two times less than the Schwarzschild radius $R_{Sch}$ of corresponding central mass. Such size and jets also predicted by FGT.

**New ideas for experimental testing the gravity physics**

From above analysis one can conclude that the question about the gravity nature will be answered soon by means of forthcoming experiments and observations, which are based on modern technological breakthroughs. Nowadays real possibility exists for testing new ideas in gravitational physics and cosmology.

In the frame of geometrical approach new observations allow to perform decisive tests of theoretical predictions for classical strong gravity effects and also the quantum space-time effects in modifications of GRT [13 − 15].

In the frame of field approach there are new ideas for gravity exploration which admit the "material agent" (gravitons) for interpretation of gravitational interaction. Possible new gravitational experiments are [10]: testing the universality of free fall for rotating bodies; search for additional scalar GW;

new principles in constructing GW detectors; new possibilities for the structure and evolution of relativistic massive objects. Possible cosmological applications of the field gravitation theory are considered in [16], which take into account global gravitational cosmological redshift and fractality of large-scale galaxy distribution.

Intriguingly, existing theoretical and observational data in the beginning of the 21$^{st}$ century point to forthcoming "paradigm shift" in understanding of the gravity physics from classical geometry to relativistic quantum theory, where Newtonian "material agent" could be Feynman's graviton which carries positive localizable energy $hv$.